\documentclass[11pt]{revtex4}   
\usepackage{amssymb,epsf}
\usepackage{latexsym}
\begin{document}
\title{Interacting new agegraphic dark energy in non-flat Brans-Dicke cosmology}
\author{Ahmad Sheykhi \footnote{sheykhi@mail.uk.ac.ir}}
\address{Department of Physics, Shahid Bahonar University, P.O. Box 76175, Kerman, Iran\\
         Research Institute for Astronomy and Astrophysics of Maragha (RIAAM), Maragha,
         Iran}

 \begin{abstract}
We construct a cosmological model of late acceleration based on
the new agegraphic dark energy model in the framework of
Brans-Dicke cosmology where the new agegraphic energy density
$\rho_{D}= 3n^2 m^2_p /\eta^{2}$ is replaced with $\rho_{D}=
{3n^2\phi^2 }/({4\omega \eta^2}$). We show that the combination of
Brans-Dicke field and agegraphic dark energy can accommodate $w_D
= -1 $ crossing for the equation of state of
\textit{noninteracting} dark energy. When an interaction between
dark energy and dark matter is taken into account, the transition
of $w_D $ to phantom regime can be more easily accounted for than
when resort to the Einstein field equations is made. In the
limiting case $\alpha = 0$ $(\omega\rightarrow \infty)$, all
previous results of the new agegraphic dark energy in Einstein
gravity are restored.
\end{abstract}

 \maketitle
 \section{Introduction\label{Int}}
One of the most dramatic discoveries of the modern cosmology in
the past decade is that our universe is currently accelerating
\cite{Rie}. A great variety of scenarios have been proposed to
explain this acceleration while most of them cannot explain all
the features of universe or they have so many parameters that
makes them difficult to fit. For a recent review on dark energy
proposals see \cite{Pad}. Many theoretical studies on the dark
energy problem are devoted to understand and shed the light on it
in the framework of a fundamental theory such as string theory or
quantum gravity. Although a complete theory of quantum gravity has
not established yet today, we still can make some attempts to
investigate the nature of dark energy according to some principles
of quantum gravity. The holographic dark energy and the agegraphic
dark energy (ADE) models are just such examples, which are
originated from some considerations of the features of the quantum
theory of gravity. That is to say, the holographic and ADE models
possess some significant features of quantum gravity. The former,
that arose a lot of enthusiasm recently \cite{Coh,wang}, is
motivated from the holographic hypothesis \cite{Suss1} and has
been tested and constrained by various astronomical observations
\cite{Xin}. The later (ADE) is based on the uncertainty relation
of quantum mechanics together with the gravitational effect in
general relativity. The ADE model assumes that the observed dark
energy comes from the spacetime and matter field fluctuations in
the universe \cite{Cai1,Wei2}. Following the line of quantum
fluctuations of spacetime, Karolyhazy et al. \cite{Kar1} discussed
that the distance $t$ in Minkowski spacetime cannot be known to a
better accuracy than $\delta{t}=\beta t_{p}^{2/3}t^{1/3}$ where
$\beta$ is a dimensionless constant of order unity. Based on
Karolyhazy relation, Maziashvili \cite{Maz} argued that the energy
density of spacetime fluctuations is given by
\begin{equation}\label{rho0}
\rho_{D} \sim \frac{1}{t_{p}^2 t^2} \sim \frac{m^2_p}{t^2},
\end{equation}
where $t_{p}$ and $m_p$ are the reduced Planck time and mass,
respectively. On these basis, Cai wrote down the energy density of
the original ADE as \cite{Cai1}
\begin{equation}\label{rhoso}
\rho_{D}=\frac{3n^2m^2_p}{T^2},
\end{equation}
where $T$ is the age of the universe and the numerical factor
$3n^2$ is introduced to parameterize some uncertainties, such as
the species of quantum fields in the universe.  However, the
original ADE model has some difficulties \cite{Cai1}. In
particular, it suffers from the difficulty to describe the
matter-dominated epoch. Therefore, a new model of ADE was proposed
\cite{Wei2}, while the time scale is chosen to be the conformal
time $\eta$ instead of the age of the universe, which is defined
by $dt= ad\eta$, where $t$ is the cosmic time. It is worth noting
that the Karolyhazy relation $\delta{t}= \beta t_{p}^{2/3}t^{1/3}$
was derived for Minkowski spacetime $ds^2 = dt^2-d\mathrm{x^2}$
\cite{Kar1,Maz}. In case of the FRW universe, we have $ds^2 =
dt^2-a^2d\mathrm{x^2} = a^2(d\eta^2-d\mathrm{x^2})$. Thus, it
might be more reasonable to choose the time scale in Eq.
(\ref{rhoso}) to be the conformal time $\eta$ since it is the
causal time in the Penrose diagram of the FRW universe. The new
ADE  contains some new features different from the original ADE
and overcome some unsatisfactory points. The ADE models have been
examined and constrained by various astronomical observations
\cite{age,shey0}.

On the other front, it is quite possible that gravity is not given
by the Einstein action, at least at sufficiently high energies. In
string theory, gravity becomes scalar-tensor in nature. The low
energy limit of string theory leads to the Einstein gravity,
coupled non-minimally to a scalar field \cite{Wit1}. Although the
pioneering study on scalar-tensor theories was done several
decades ago \cite{BD}, it has got a new impetus recently as it
arises naturally as the low energy limit of many theories of
quantum gravity such as superstring theory or Kaluza-Klein theory.
Because the agegraphic energy density belongs to a dynamical
cosmological constant, we need a dynamical frame to accommodate it
instead of Einstein gravity. Therefore the investigation on the
agegraphic models of dark energy in the framework of Brans-Dicke
theory is well motivated. In the framework of Brans-Dicke
cosmology, holographic models of dark energy have also been
studied \cite{Pavon2}. Our aim in this paper is to construct a
cosmological model of late acceleration based on the Brans-Dicke
theory of gravity and on the assumption that the pressureless dark
matter and new ADE do not conserve separately but interact with
each other.
\section{NEW ADE in Branse-Dicke Theory\label{NEW}}
We start from the action of Brans-Dicke theory which in the
canonical form can be written \cite{Arik}
\begin{equation}
 S=\int{
d^{4}x\sqrt{g}\left(-\frac{1}{8\omega}\phi ^2
{R}+\frac{1}{2}g^{\mu \nu}\partial_{\mu}\phi \partial_{\nu}\phi
+L_M \right)},\label{act1}
\end{equation}
where ${R}$ is the scalar curvature and $\phi$ is the Brans-Dicke
scalar field. The non-minimal coupling term $\phi^2 R$ replaces
with the Einstein-Hilbert term ${R}/{G}$ in such a way that
$G^{-1}_{\mathrm{eff}}={2\pi \phi^2}/{\omega}$  where
$G_{\mathrm{eff}}$ is the effective gravitational constant as long
as the dynamical scalar field $\phi$ varies slowly. The signs of
the non-minimal coupling term and the kinetic energy term are
properly adopted to $(+---)$ metric signature. The new ADE model
will be accommodated in the non-flat Friedmann-Robertson-Walker
(FRW) universe which is described by the line element
\begin{eqnarray}
 ds^2=dt^2-a^2(t)\left(\frac{dr^2}{1-kr^2}+r^2d\Omega^2\right),\label{metric}
 \end{eqnarray}
where $a(t)$ is the scale factor, and $k$ is the curvature
parameter with $k = -1, 0, 1$ corresponding to open, flat, and
closed universes, respectively. A closed universe with a small
positive curvature ($\Omega_k\simeq0.01$) is compatible with
observations \cite{spe}. Varying action (\ref{act1}) with respect
to metric (\ref{metric}) for a universe filled with dust and ADE
yields the following field equations
\begin{eqnarray}
 &&\frac{3}{4\omega}\phi^2\left(H^2+\frac{k}{a^2}\right)-\frac{1}{2}\dot{\phi} ^2+\frac{3}{2\omega}H
 \dot{\phi}\phi=\rho_m+\rho_D,\label{FE1} \\
&&\frac{-1}{4\omega}\phi^2\left(2\frac{{\ddot{a}}}{a}+H^2+\frac{k}{a^2}\right)-\frac{1}{\omega}H
\dot{\phi}\phi-\frac{1}{2\omega}
 \ddot{\phi}\phi
-\frac{1}{2}\left(1+\frac{1}{\omega}\right)\dot{\phi}^2=p_D,\label{FE2}\\
&&\ddot{\phi}+3H
 \dot{\phi}-\frac{3}{2\omega}\left(\frac{{\ddot{a}}}{a}+H^2+\frac{k}{a^2}\right)\phi=0,
 \label{FE3}
\end{eqnarray}
where the dot is the derivative with respect to time and
$H=\dot{a}/a$ is the Hubble parameter. Here $\rho_D$, $p_D$ and
$\rho_m$ are, respectively, the dark energy density, dark energy
pressure and energy density of dust (dark matter). We shall assume
that Brans-Dicke field can be described as a power law of the
scale factor, $\phi\propto a^{\alpha}$. A case of particular
interest is that when $\alpha$ is small whereas $\omega$ is high
so that the product $\alpha \omega$ results of order unity
\cite{Pavon2}. This is interesting because local astronomical
experiments set a very high lower bound on $\omega$; in
particular, the Cassini experiment implies that $\omega>10^4$
\cite{Bert}. Taking the derivative with respect to time of
relation $\phi\propto a^{\alpha}$, we get
\begin{eqnarray}\label{dotphi}
&&\dot{\phi}=\alpha H \phi, \\
&&\ddot{\phi}=\alpha^2 H^2\phi+\alpha\phi\dot{H}.\label{ddotphi}
\end{eqnarray}
The energy density of the new ADE can be written \cite{Wei2}
\begin{equation}\label{rhosn}
\rho_{D}= \frac{3n^2 m_{p}^2}{\eta^2},
\end{equation}
where the conformal time is given by
\begin{equation}
\eta=\int_0^a{\frac{da}{Ha^2}}.
\end{equation}
In the framework of Brans-Dicke cosmology, we write down the new
agegraphic energy density of the quantum fluctuations in the
universe as
\begin{equation}\label{rho1n}
\rho_{D}= \frac{3n^2\phi^2 }{4\omega \eta^2}.
\end{equation}
where $\phi^2={\omega}/{2\pi G_{\mathrm{eff}}}$. In the limit of
Einstein gravity, $G_{\mathrm{eff}}\rightarrow G$, expression
(\ref{rho1n}) recovers the standard new agegraphic energy density
in Einstein gravity. We define the critical energy density,
$\rho_{\mathrm{cr}}$, and the energy density of the curvature,
$\rho_k$, as
\begin{eqnarray}\label{rhocr}
\rho_{\mathrm{cr}}=\frac{3\phi^2 H^2}{4\omega},\hspace{0.8cm}
\rho_k=\frac{3k\phi^2}{4\omega a^2}.
\end{eqnarray}
We also introduce, as usual, the fractional energy densities such
as
\begin{eqnarray}
&&\Omega_m=\frac{\rho_m}{\rho_{\mathrm{cr}}}=\frac{4\omega\rho_m}{3\phi^2
H^2}, \label{Omegam} \\
&&\Omega_k=\frac{\rho_k}{\rho_{\mathrm{cr}}}=\frac{k}{H^2
a^2}\label{Omegak}\\
&&\Omega_D=\frac{\rho_D}{\rho_{\mathrm{cr}}}=\frac{n^2}{H^2\eta^2}
\label{OmegaDn}.
\end{eqnarray}
\subsection{Noninteracting case}
Let us begin with the noninteracting case, in which the dark
energy and dark matter evolves according to their conservation
laws
\begin{eqnarray}
&&\dot{\rho}_D+3H\rho_D(1+w_D)=0,\label{consq}\\
&&\dot{\rho}_m+3H\rho_m=0, \label{consm}
\end{eqnarray}
where $w_D=p_D/\rho_D$ is the equation of state parameter of the
new ADE. Differentiating Eq. (\ref{rho1n}) and using Eqs.
(\ref{dotphi}) and (\ref{OmegaDn}) we have
\begin{eqnarray}
\dot{\rho}_D=2H\rho_D\left(\alpha-\frac{\sqrt{\Omega_D}}{na}\right)\label{rhodotn}.
\end{eqnarray}
Inserting this equation in the conservation law (\ref{consq}), we
obtain the equation of state parameter of the new ADE
\begin{eqnarray}
w_D=-1-\frac{2\alpha}{3}+\frac{2}{3na}\sqrt{\Omega_D}\label{wDn}.
\end{eqnarray}
It is important to note that when  $\alpha=0$, the Brans-Dicke
scalar field becomes trivial and Eq. (\ref{wDn}) reduces to its
respective expression in new ADE in general relativity \cite{Wei2}
\begin{eqnarray}
w_D=-1+\frac{2}{3na}\sqrt{\Omega_D}\label{wDnstand}.
\end{eqnarray}
In this case ($\alpha=0$), the present accelerated expansion of
our universe can be derived only if $n>1$ \cite{Wei2}. Note that
we take $a=1$ for the present time. In addition, $w_D$ is always
larger than $-1$ and cannot cross the phantom divide $w_D =-1$.
However, in the presence of the Brans-Dicke field  ($\alpha>0$)
the condition $n>1$ is no longer necessary to derive the present
accelerated expansion. Besides, from Eq. (\ref{wDn}) one can
easily see that $w_D$ can cross the phantom divide provided
$na\alpha>\sqrt{\Omega_D}$. If we take $\Omega_D=0.73$ and $a=1$
for the present time, the phantom-like equation of state can be
accounted if $n\alpha>0.85$. For instance, for $n=1$ and
$\alpha=0.9$, we get $w_D=-1.03$. Therefore, with the combination
of new agegraphic energy density with the Brans-Dicke field $w_D$
of \textit{noninteracting} new ADE can cross the phantom divide.

Let us examine the behavior of $w_D$ in two different stages. In
the late time where $\Omega_D \rightarrow 1$ and $a
\rightarrow\infty$ we have $w_D=-1-\frac{2\alpha}{3}$. Thus
$w_D<-1$ for $\alpha>0$. This implies that in the late time $w_D$
necessary crosses the phantom divide in the framework of
Brans-Dicke theory. In the early time where $\Omega_D \rightarrow
0$  and $a\rightarrow 0$ we cannot find $w_D$ from Eq. (\ref{wDn})
directly. Let us consider the matter-dominated epoch,
$H^2\propto\rho_m\propto a^{-3}$. Therefore $\sqrt{a}da\propto
dt=ad\eta$. Thus $\eta\propto \sqrt{a}$. From Eq. (\ref{rho1n}) we
have $\rho_D\propto a^{2\alpha-1}$. Putting this in conservation
law, $\dot{\rho}_D+3H\rho_D(1+w_D)=0$, we obtain
$w_D=-{2}/{3}-2\alpha/3$. Substituting this $w_D$ in Eq.
(\ref{wDn}) we find that $\Omega_D=n^2a^2/4$ in the matter
dominated epoch as expected. We will see below that this is
exactly the result one obtains for $\Omega_D$ from its equation of
motion in the matter-dominated epoch.

Since in our model the dynamics of the scale factor is governed
not only by the dark matter and new ADE, but also by the
Brans-Dicke field, the signature of the deceleration parameter,
$q=-\ddot{a}/(aH^2)$, has to be examined carefully. When the
deceleration parameter is combined with the Hubble parameter and
the dimensionless density parameters, form a set of useful
parameters for the description of the astrophysical observations.
Dividing  Eq. (\ref{FE2}) by $H^2$, and using Eqs. (\ref{dotphi}),
(\ref{ddotphi}) and (\ref{rho1n})-(\ref{OmegaDn}) we obtain
\begin{eqnarray}
q=\frac{1}{2\alpha+2}\left[(2\alpha+1)^2+2\alpha(\alpha\omega-1)+\Omega_k+3\Omega_D
w_D\right]\label{q1}.
\end{eqnarray}
Substituting $w_D$ from Eq. (\ref{wDn}), we reach
\begin{eqnarray}
q&=&\frac{1}{2\alpha+2}\left[(2\alpha+1)^2+2\alpha(\alpha\omega-1)+\Omega_k
-(2\alpha+3)\Omega_D+\frac{2}{na}{\Omega^{3/2}_D}\right]\label{q2n}.
\end{eqnarray}
When $\alpha=0$, Eq. (\ref{q2n}) restores the deceleration
parameter of the new ADE in general relativity \cite{shey0}
\begin{eqnarray}
q=\frac{1}{2}(1+\Omega_k)-\frac{3}{2}\Omega_D+\frac{\Omega^{3/2}_D}{na}\label{q3n}.
\end{eqnarray}
Finally, we obtain the equation of motion for $\Omega_D$. Taking
the derivative of Eq. (\ref{OmegaDn}) and using relation
${\dot{\Omega}_D}=H{\Omega'_D}$, we get
\begin{eqnarray}\label{OmegaD2n}
{\Omega'_D}=\Omega_D\left(-2\frac{\dot{H}}{H^2}-\frac{2}{na
}\sqrt{\Omega_D}\right),
\end{eqnarray}
where the prime denotes the derivative with respect to $x=\ln{a}$.
Using relation $q=-1-\frac{\dot{H}}{H^2}$, we have
\begin{eqnarray}\label{OmegaD3n}
{\Omega'_D}=2\Omega_D\left(1+q-\frac{\sqrt{\Omega_D}}{na}\right),
\end{eqnarray}
where $q$ is given by Eq. (\ref{q2n}). Let us examine the above
equation for matter dominated epoch where $a\ll1$ and $\Omega_D
\ll1$. Substituting $q$ from Eq. (\ref{q2n}) in (\ref{OmegaD3n})
with $\Omega_k\ll1$, $\alpha\ll1$  and $\alpha \omega\approx 1$,
this equation reads as
\begin{eqnarray}\label{OmegaD4n}
\frac{d\Omega_D}{da}\simeq\frac{\Omega_D}{a}
\left(3-\frac{2}{na}\sqrt{\Omega_D}\right).
\end{eqnarray}
Solving this equation we find $\Omega_D =n^2 a^2/4$, which is
consistent with our previous result. Therefore, all things are
consistent. The confusion in the original ADE is removed in this
new model.
\subsection{Interacting case}
Next we generalize our study to the case where the pressureless
dark matter and the new ADE do not conserve separately but
interact with each other. Given the unknown nature of both dark
matter and dark energy there is nothing in principle against their
mutual interaction and it seems very special that these two major
components in the universe are entirely independent. Indeed, this
possibility is receiving growing attention in the literature
\cite{Ame} and appears to be compatible with SNIa and CMB data
\cite{Oli}. The total energy density satisfies a conservation law
\begin{equation}\label{cons}
\dot{\rho}+3H(\rho+p)=0.
\end{equation}
However, since we consider the interaction between dark matter and
dark energy, $\rho_{m}$ and $\rho_{D}$ do not conserve separately;
they must rather enter the energy balances
\begin{eqnarray}
&&\dot{\rho}_m+3H\rho_m=Q, \label{consmI}
\\&& \dot{\rho}_D+3H\rho_D(1+w_D)=-Q,\label{consqI}
\end{eqnarray}
where $Q =\Gamma\rho_D$ stands for the interaction term with
$\Gamma>0$. Using Eqs.  (\ref{dotphi}) and (\ref{rhocr}), we can
rewrite the first Friedmann equation (\ref{FE1}) as
\begin{eqnarray}\label{rhos}
\rho_{\mathrm{cr}}+\rho_k=\rho_m+\rho_D+\rho_{\phi},
\end{eqnarray}
where we have defined
\begin{eqnarray}\label{rhophi}
\rho_{\phi}\equiv\frac{1}{2}\alpha
H^2\phi^2\left(\alpha-\frac{3}{\omega}\right).
\end{eqnarray}
Dividing Eq. (\ref{rhos}) by  $\rho_{\mathrm{cr}}$, this equation
can be written as
\begin{eqnarray}\label{Fried2new}
\Omega_m+\Omega_D+\Omega_{\phi}=1+\Omega_k,
\end{eqnarray}
where
\begin{eqnarray}\label{Omegaphi}
\Omega_{\phi}=\frac{\rho_{\phi}}{\rho_{\mathrm{cr}}}=-2\alpha\left(1-\frac{\alpha\omega}{3}\right).
\end{eqnarray}
We also assume $\Gamma=3b^2(1+r)H$ where $r={\rho_m}/{\rho_D}$ and
$b^2$ is a coupling constant. Therefore, the interaction term $Q$
can be expressed as
\begin{eqnarray}\label{Q}
Q=3b^2H\rho_D(1+r),
\end{eqnarray}
where
\begin{eqnarray}\label{r}
r&=&\frac{\Omega_m}{\Omega_D}
=-1+{\Omega^{-1}_D}\left[1+\Omega_k+2\alpha\left(1-\frac{\alpha\omega}{3}\right)\right].
\end{eqnarray}
Combining Eqs. (\ref{rhodotn}), (\ref{Q}) and (\ref{r}) with  Eq.
(\ref{consqI}) we can obtain the equation of state parameter
\begin{eqnarray}
w_D&=&-1-\frac{2\alpha}{3}+\frac{2}{3na}\sqrt{\Omega_D}-b^2
{\Omega^{-1}_D}\left[1+\Omega_k+2\alpha\left(1-\frac{\alpha\omega}{3}\right)\right]\label{wDnInt}.
\end{eqnarray}
When  $\alpha=0$, Eq. (\ref{wDnInt}) recovers its respective
expression of interacting new ADE model in general relativity
\cite{shey0}. From Eq. (\ref{wDnInt}) we see that with the
combination of the new ADE and Brans-Dicke field, the transition
of $w_D $ from the phantom divide can be more easily accounted
than in Einstein gravity. For completeness we also present the
deceleration parameter for the interacting case
\begin{eqnarray}
q&=&\frac{1}{2\alpha+2}\left[(2\alpha+1)^2+2\alpha(\alpha\omega-1)+\Omega_k-(2\alpha+3)\Omega_D
+\frac{2}{na}{\Omega^{3/2}_D}\right. \nonumber\
\\
&&
\left.-3b^2\left(1+\Omega_k+2\alpha\left(1-\frac{\alpha\omega}{3}\right)\right)\right]\label{q2nInt}.
\end{eqnarray}
In the limiting case  $\alpha=0$, Eq. (\ref{q2nInt}) restores the
deceleration parameter for the standard interacting new ADE in a
non-flat universe \cite{shey0}
\begin{eqnarray}
q&=&\frac{1}{2}(1+\Omega_k)-\frac{3}{2}{\Omega_D}
+\frac{\Omega^{3/2}_D}{na}-\frac{3b^2}{2}(1+\Omega_k)\label{q3Int}.
\end{eqnarray}
For flat universe, $\Omega_k=0$,  and we recover exactly the
result of \cite{Wei2}. The equation of motion for $\Omega_D$ takes
the form (\ref{OmegaD3n}), where $q$ is now given by Eq.
(\ref{q2nInt}).
\section{Conclusions\label{CONC}}
An interesting attempt for probing the nature of dark energy
within the framework of quantum gravity is the so-called ADE
proposal. Since ADE models belong to a dynamical cosmological
constant, it is more natural to study them in the framework of
Brans-Dicke theory than in Einstein gravity. In this paper, we
studied a cosmological model of late acceleration based on the new
ADE model in the framework of non-flat Brans-Dicke cosmology where
the new agegraphic energy density $\rho_{D}= {3n^2
m^2_p}/\eta^{2}$ is replaced with $\rho_{D}= {3n^2\phi^2
}/({4\omega \eta^2})$. With this replacement in Brans-Dicke
theory, we found that the acceleration of the universe expansion
will be more easily achieved for than when the standard new ADE in
general relativity is employed. Interestingly enough, we found
that with the combination of Brans-Dicke field and ADE the
equation of state of \textit{noninteracting} new ADE can cross the
phantom divide. This is in contrast to Einstein gravity where the
equation of state of \textit{noninteracting} new ADE cannot cross
the phantom divide \cite{Wei2}. When an interaction between dark
energy and dark matter is taken into account, the transition to
phantom regime for the equation of state of new ADE can be more
easily accounted for than when resort to the Einstein field
equations is made.

\acknowledgments{I thank the anonymous referee for constructive
comments. This work has been supported by Research Institute for
Astronomy and Astrophysics of Maragha, Iran.}


\begin{thebibliography}{99}

\bibitem{Rie} A.G. Riess, et al., Astron. J.  116 (1998)
1009; S. Perlmutter, et al.,  Astrophys. J.  517 (1999) 565; P. de
Bernardis, et al.,  Nature  404 (2000) 955.

\bibitem{Pad} T. Padmanabhan, Phys. Rep.  {{380}},  (2003) 235; E.J. Copeland, M. Sami, S. Tsujikawa, Int. J. Mod. Phys. D
15 (2006) 1753.


\bibitem{Coh}  A. Cohen, D. Kaplan, A. Nelson, Phys. Rev. Lett. 82 (1999)
4971; M. Li, Phys. Lett. B 603 (2004) 1; Q. G. Huang, M. Li, JCAP
0408 (2004) 013; S.D. H. Hsu, Phys. Lett. B 594 (2004) 13.



\bibitem{wang}D. Pavon, W. Zimdahl, Phys. Lett. B 628 (2005) 206;
 B. Wang, Y. Gong and E. Abdalla, Phys. Lett. B 624
(2005) 141; B. Wang, C. Y. Lin and E. Abdalla, Phys. Lett. B 637
(2005) 357; B. Wang, C. Y. Lin. D. Pavon and E. Abdalla, Phys.
Lett. B 662 (2008) 1; M. R. Setare, S. Shafei, JCAP 09 (2006) 011;
M. R. Setare, E. C. Vagenas, Phys. Lett. B 666 (2008) 111.

\bibitem{Suss1}  G. 't Hooft, gr-qc/9310026; L. Susskind, J. Math. Phys. 36 (1995)
6377.

\bibitem{Xin} X. Zhang, F. Q. Wu,  Phys. Rev. D 72 (2005)
043524; X. Zhang, F. Q.  Wu, Phys. Rev. D 76 (2007) 023502; B.
Feng, X. Wang, X. Zhang, Phys. Lett. B 607 (2005) 35.



\bibitem{Cai1} R. G. Cai, Phys. Lett. B 657 (2007) 228.


\bibitem{Wei2} H. Wei and R. G. Cai, Phys. Lett. B 660 (2008) 113.

\bibitem{Kar1} F. Karolyhazy,
Nuovo.Cim. A 42 (1966) 390; F. Karolyhazy, A. Frenkel and B.
Lukacs, in \textit{Physics as natural Philosophy}  edited by A.
Shimony and H. Feschbach, MIT Press, Cambridge, MA, (1982); F.
Karolyhazy, A. Frenkel and B. Lukacs, in \textit{Quantum Concepts
in Space and Time}  edited by R. Penrose and C.J. Isham, Clarendon
Press, Oxford, (1986).




\bibitem{Maz} M. Maziashvili, Int. J. Mod. Phys. D 16 (2007) 1531; M. Maziashvili, Phys. Lett. B 652 (2007) 165.


\bibitem{age} H. Wei and R. G. Cai, Eur. Phys. J. C 59 (2009) 99; H. Wei and R. G. Cai, Phys. Lett. B 663 (2008) 1; K. Y. Kim, H. W. Lee, Y. S. Myung, Phys.Lett. B 660 (2008)
118.
 \bibitem{shey0}  A. Sheykhi, Phys. Lett. B 680 (2009) 113. 



\bibitem{Wit1}  M. B. Green, J. H. Schwarz and E. Witten, {\it Superstring
Theory}, Cambridge University Press, Cambridge (1987).

\bibitem{BD} C. Brans and R. H. Dicke, Phys. Rev. 124 (1961) 925.

\bibitem{Pavon2} N. Banerjee, D. Pavon, Phys. Lett. B 647 (2007) 447; M. R.  Setare, Phys. Lett. B 644 (2007)
99; A. Sheykhi, Phys.  Lett.  B 681 (2009) 205.

\bibitem{Arik}  M. Arik, M.C. Calik, Mod. Phys. Lett. A  21 (2006)
1241.

\bibitem{spe} D. N. Spergel, Astrophys. J. Suppl. 148 (2003) 175; C. L. Bennett, et al.,  Astrophys. J. Suppl.
148 (2003) 1.


\bibitem{Bert} B. Bertotti, L. Iess and P. Tortora, Nature, 425 (2003)
374; V. Acquaviva, L. Verde, JCAP 12 (2007) 001.


\bibitem{Ame} L. Amendola, Phys. Rev. D 60 (1999)  043501; L. Amendola, Phys. Rev. D 62 (2000) 043511;
 L. Amendola and C. Quercellini, Phys. Rev. D 68
(2003)  023514; W. Zimdahl and D. Pavon, Phys. Lett. B 521 (2001)
133; W. Zimdahl and D. Pavon, Gen. Rel. Grav. 35 (2003) 413.


\bibitem{Oli} G. Olivares, F. Atrio, D. Pavon, Phys. Rev. D 71 (2005) 063523.




\end{thebibliography}
\end{document}